\documentclass[twocolumn,letter]{jpsj3}
\usepackage{color}
\title{Trimer Formation and Metal-Insulator Transition in \\
Orbital Degenerate Systems on a Triangular Lattice}

\author{Junki \surname{Yoshitake}\thanks{E-mail address: yoshitake@aion.t.u-tokyo.ac.jp} and \name{Yukitoshi \surname{Motome}}
}
\inst{\address{Department of Applied Physics, University of Tokyo, 7-3-1 Hongo, Bunkyo, Tokyo 113-8656, Japan}
}
\abst{As a prototypical self-organization in the system with orbital degeneracy, we theoretically investigate 
trimer formation on a triangular lattice, as observed in LiVO$_2$.
From the analysis of an effective spin-orbital coupled model in the strong correlation limit, 
we show that the previously-proposed orbital-ordered trimer state is not the lowest-energy state 
for a finite Hund's-rule coupling. 
Instead, 
exploring the ground state in a wide range of parameters for a multiorbital Hubbard model, 
we find an instability toward  
a different orbital-ordered trimer state 
in the intermediately correlated regime 
in the presence of trigonal crystal field. 
The trimer phase appears 
in the competing region among a paramagnetic metal, band insulator, and Mott insulator. 
The underlying mechanism is nesting instability of the Fermi surface 
by a synergetic effect of Coulomb interactions and trigonal-field splitting. 
The results are compared with experiments in 
triangular-lattice compounds, 
LiV$X_2$ ($X$=O, S, Se) and NaVO$_2$.}

\kword{cluster formation, trimer, multiorbital Hubbard model, triangular lattice, 
orbital order, metal-insulator transition}

\begin{document}
\maketitle

Orbital degree of freedom plays a decisive role in many transition metal compounds~\cite{Kugel1982,Tokura2000,Khomskii2006}.
Orbital ordering and its fluctuation affect 
magnetic, transport, and structural properties in a complicated manner 
through the coupling to spin, charge, and lattice degrees of freedom. 
One of the characteristic phenomena, 
particularly observed in geometrically-frustrated systems, 
is spontaneous formation of clusters. 
It is a structural phase transition 
to form `molecules' of transition metal atoms by periodic modulation of the lattice structure.
For example, CuIr$_2$S$_4$ shows a formation of 
octumers~\cite{Radaelli2002} and AlV$_2$O$_4$ shows heptamers~\cite{Horibe2006}
on their pyrochlore networks of Ir or V, and 
the importance of $d$-electron orbitals was theoretically pointed out~\cite{Khomskii2005,Matsuda2006}.  
The spontaneous formation of `molecules' has attracted increasing interest 
as emergent physics in the systems with multiple degrees of freedom.

A prototypical example of 
such clusters is a trimer 
in a quasi-two-dimensional compound 
LiVO$_2$~\cite{Bongers1957,Goodenough}. 
In this system, each V$^{3+}$ cation has two $3d$ electrons in threefold $t_{2g}$ orbitals on average, 
and forms triangular lattices.
The compound exhibits a first-order
structural phase transition at $T_{\rm{c}} \simeq 500$~K, 
accompanied by a sudden drop of the magnetic susceptibility~\cite{Bongers1957,Onoda1991,Tian2004}.
Below $T_{\rm{c}}$, 
V-V bonds are periodically modulated in each layer 
to form V$_3$ trimers~\cite{Goodenough,Cardoso1988,Imai1995,Tian2004}. 
The trimer formation was first interpreted by a charge-density-wave instability 
associated with the V clustering~\cite{Goodenough,Goodenough1991}. 
A spin-Peierls type scenario was also considered to account for the spin-singlet nature 
below $T_{\rm c}$~\cite{Onoda1991}. 
Later, Pen {\it et al.} pointed out a crucial role of orbital ordering~\cite{Pen1997}: 
It was claimed that LiVO$_2$ is a Mott insulator 
and the trimers are stabilized as the $S=1$ spin-singlet objects under 
a particular ordering of the $t_{2g}$ orbitals. 

Experimentally, however, it is not clear to what extent the electron correlation plays 
an important role in LiVO$_2$. 
The spin gap in the low-temperature($T$) trimer phase was estimated to be 
$\Delta_{\rm s} \sim 1600$~K~\cite{Onoda1991}, 
which is comparable to the charge gap estimated from $T$ dependence of the resistivity, 
$\Delta_{\rm c} \sim 0.14$~eV~\cite{Tian2004}. 
In addition, the Curie-Weiss temperature 
estimated from the magnetic susceptibility 
for $T>T_{\rm c}$ is $\sim -1500$~K~\cite{Tian2004}, 
whose magnitude is also comparable to $\Delta_{\rm c}$. 
These facts clearly contradict with the local moment picture under the strong correlation. 
Recently, it was also shown that the substitution of O by S or Se 
makes the system metallic~\cite{Katayama2009}. 
This suggests that LiVO$_2$ is located 
in the vicinity of correlation-driven metal-insulator transition. 
It is highly nontrivial how the itinerant tendency is compatible with 
the $S=1$ local moment formation expected for the strong electron correlation. 
Hence, the origin of the trimer formation in LiVO$_2$ is still in dispute. 
It is desired to revisit the problem theoretically 
for clarifying the role of electron correlation and orbital degree of freedom. 

In this Letter, we address this issue by carefully examining 
various orbital and spin orderings in a wide range of electron correlation. 
In particular, revisiting the strong correlation picture, 
we show that the system is likely to exhibit a 
four-sublattice spin and orbital ordering in the ground state, 
not the previously-proposed trimer state. 
We, however, identify a different orbital-ordered trimer state in the intermediately correlated region 
on the verge of metal-insulator transition, 
where the Coulomb interactions work cooperatively with the trigonal-field splitting.
The results provide underlying mechanism in the self-organization 
in orbital degenerate systems close to metal-insulator transition.

We start with a multiorbital Hubbard model for the threefold $t_{2g}$ orbitals.
The Hamiltonian is given as
\begin{align}
\mathcal{H}
=  &- \sum_{\left\langle{ij}\right\rangle}\sum_{\alpha\beta}\sum_{\tau} t^{\alpha \beta}_{ij} \left( c^{\dagger}_{i\alpha\tau} c_{j\beta\tau} + \rm{H.c.} \right) \notag\\
 & 
+ \frac12 \sum_i \sum_{\alpha \beta \alpha^\prime \beta^\prime} \sum_{\tau \tau^\prime}
U_{\alpha \beta \alpha^\prime \beta^\prime}
c_{i\alpha \tau}^\dagger c_{i\beta \tau^\prime}^\dagger 
c_{i\beta^\prime \tau^\prime} c_{i\alpha^\prime \tau},
\label{eq:fullHamiltonian0}
\end{align}
where the first term is the electron hopping 
between the nearest-neighbor sites $\langle ij \rangle$ on a triangular lattice; 
$\alpha, \beta$ denote the three $t_{2g}$ orbitals, $d_{xy}$, $d_{yz}$, or $d_{zx}$; 
$\tau,\tau^\prime$ denote the spins, $\uparrow$ or $\downarrow$. 
The second term represents the onsite Coulomb interactions, for which 
we use the standard parametrizations, 
$
U_{\alpha \beta \alpha^\prime \beta^\prime} 
= U^\prime \delta_{\alpha \alpha^\prime} \delta_{\beta \beta^\prime} 
+ J_{\rm H} (\delta_{\alpha \beta^\prime} \delta_{\beta \alpha^\prime} 
+ \delta_{\alpha \beta} \delta_{\alpha^\prime \beta^\prime})
$
and 
$
U = U^\prime + 2 J_{\rm H}
$.
Following the previous study~\cite{Pen1997},  
we here consider the overlap integrals of $\sigma$-bond orbitals only and 
take them as the energy unit, $t_\sigma = 1$. 
We fix the electron density at 
$n=\sum_{i\alpha} \langle n_{i\alpha} \rangle /N =2$, i.e., 
two electrons per site on average ($N$ is the number of sites). 

First, we consider the strong correlation limit of the model in Eq.~(\ref{eq:fullHamiltonian0}). 
The second order perturbation in $t_\sigma/U$ gives an effective spin-orbital Hamiltonian~\cite{Kugel1982} 
in the form
\begin{align}
\mathcal{H}_{\rm{eff}}
=-J\sum_{\langle ij\rangle}[h_{\rm{o-AF}}^{(ij)}+h_{\rm{o-F}}^{(ij)}], \label{eq:Heff}
\end{align}
where
$
h_{\rm{o-AF}}^{(ij)} 
=(A+B\vec{S}_{i}\cdot \vec{S}_{j})(n_{i\alpha (ij)}\bar{n}_{j\alpha (ij)}+\bar{n}_{i\alpha (ij)}n_{j\alpha (ij)})
$ and
$
h_{\rm{o-F}}^{(ij)} 
=C(1-\vec{S}_{i}\cdot \vec{S}_{j})n_{i\alpha (ij)}n_{j\alpha (ij)} 
$.
Here $\vec{S}_i$ is the $S=1$ spin operator at site $i$, 
$\alpha(ij)$ is the orbital which has the $\sigma$-bond overlap 
between the sites $i$ and $j$, and $\bar{n}_{i\alpha (ij)} = 1 - n_{i\alpha (ij)}$. 
The coefficients are given as $J=t_\sigma^2/U$, 
$A=(1-\eta )/(1-3\eta )$, $B=\eta /(1-3\eta )$, $C=(1+\eta )/(1+2\eta )$, 
and $\eta =J_{\rm{H}}/U$.
The same effective model was studied for LiVO$_2$~\cite{Pen1997}: 
It was shown that an $S=1$ spin-singlet 
state with trimer-type orbital ordering [Fig.~1(a)] has 
the same ground-state energy as an antiferromagnetic (AF) state 
with a `square-type' orbital ordering [Fig.~1(b)]  
in the limit of $\eta=0$, i.e., $U=U^\prime$ and $J_{\rm H}=0$. 
However, it was not clarified how a finite Hund's-rule coupling modifies the results, 
and moreover, whether there are other competing states.

\begin{figure}[t]
\begin{center}
\includegraphics[width=7.5cm,clip]{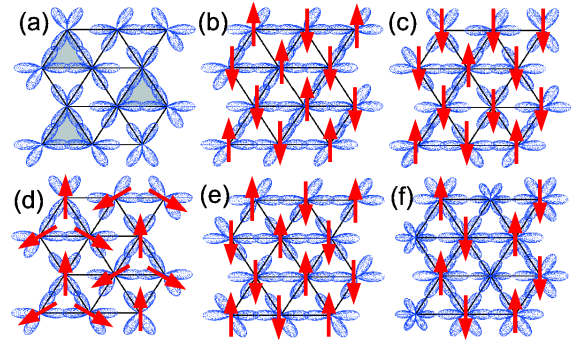}
\caption{(Color online). 
Schematic pictures of the spin-orbital ordered states: 
(a) orbital trimer state proposed in Ref.~\citen{Pen1997}, 
(b) AF state with square-type orbital ordering, 
(c) ferrimagnetic state with four-sublattice spin-orbital ordering, 
(d)-(f) Hartree-Fock solutions appearing in the phase diagram in Fig.~\ref{fig:sankakusouzu24}(a).
Only dominant orbitals are drawn by lobes in each figure. Arrows represent spins. 
In (a), the shaded triangles denote the trimers with spin-singlet formation. 
}
\label{fig:mosikizu}
\end{center}
\end{figure}

In order to search the ground state of the model in Eq.~(\ref{eq:Heff}) in an unbiased way, 
we performed Monte Carlo (MC) simulation at low $T$ for the classical counterpart of the model, 
namely, by replacing the $S=1$ operators by the classical vectors with unit length.
The results indicate that, for $\eta>0$, 
a four-sublattice ferrimagnetic state shown in Fig.~\ref{fig:mosikizu}(c) is selected as 
the lowest-energy state as $T \to 0$.

\begin{figure}[t]
\begin{center}
\includegraphics[width=7.5cm,clip]{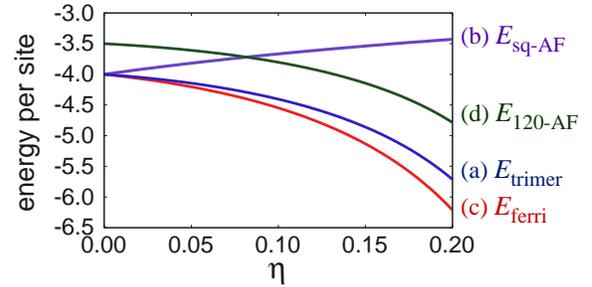}
\caption{(Color online). 
Ground-state energy of the model (\ref{eq:Heff}) 
for the four states in Figs.~\ref{fig:mosikizu}(a)-(d) 
as a function of $\eta = J_{\rm H}/U$. 
See text for details.
}
\label{fig:MonteCarloResult}
\end{center}
\end{figure}

On the basis of the MC result, 
we compare the ground-state energy of the ferrimagnetic state with other
several typical states, as plotted in Fig.~\ref{fig:MonteCarloResult}. 
Here we consider four states in Figs.~\ref{fig:mosikizu}(a)-(d). 
For the states (b)-(d), 
the ground-state energies are analytically obtained as 
$E_{\rm sq-AF} = -4CJ$,
$E_{\rm ferri} = -(2A+B+2C)J$,
$E_{\rm 120-AF} = -(2A-B+3C/2)J$, 
respectively, by treating the model (\ref{eq:Heff}) at the classical level.
For the trimer state (a), 
assuming the $S=1$ spin-singlet state in each trimer, 
namely, $\langle \vec{S}_i \cdot \vec{S}_j \rangle = -1$ for the intra-trimer bonds, 
and neglecting the inter-trimer spin correlations, we obtain 
$E_{\rm trimer} = -(2A+2C)J$. 
The comparison indicates that the four-sublattice ferrimagnetic state gives 
the lowest energy for $\eta>0$, consistent with the MC search. 
The energy is even lower than the previously-proposed trimer state~\cite{note3}. 
$E_{\rm ferri}$ will be lowered when considering 
quantum fluctuations beyond the classical level, such as the spin wave contribution. 
Therefore, our results strongly suggest that 
the trimer state proposed in Ref.~\cite{Pen1997} is not the ground state 
in the strong correlation limit, 
and is likely taken over by the four-sublattice ferrimagnetic state in Fig.~\ref{fig:mosikizu}(c).

The above considerations in the large-$U$ limit lead us to
go back to the multiorbital Hubbard model given by Eq.~(\ref{eq:fullHamiltonian0}) 
and explore another possibility 
for the trimer formation. 
In the following analysis, we extend the model by including the
trigonal distortion of VO$_6$ octahedra which is
inherent in the layered materials. 
That is, we consider an additional term to Eq.~(\ref{eq:fullHamiltonian0}) given by
\begin{equation}
\mathcal{H}_D = 
\frac{D}{2}\sum_{i}\sum_{\alpha\ne\beta}\sum_{\tau}c^{\dagger}_{i \alpha \tau}c_{i \beta \tau},
\end{equation}
which splits the $t_{2g}$ levels into $a_{1g}$ singlet and $e_g^\prime$ doublet by $3D/2$. 
The sign and magnitude of $D$ is strongly dependent on 
the detailed band structure including the $t_{2g}$-$e_g$ hybridization~\cite{Landron2008}.
In the following calculations, 
we consider the case of $D>0$ which lowers the $a_{1g}$ level~\cite{Ezhov1998}. 

To map out the ground-state phase diagram 
in a wide parameter region, we employ the Hartree-Fock approximation. 
In the calculations, we take the 
12-site unit cell in the form shown in Fig.~\ref{fig:mosikizu}, 
and consider $24\times 24$ array of the unit cell with appropriate boundary conditions. 
This enables to incorporate
both three- 
and four-sublattice orders,
such as the trimer and ferrimagnetic states.

\begin{figure}[t]
\begin{center}
\includegraphics[width=7.5cm,clip]{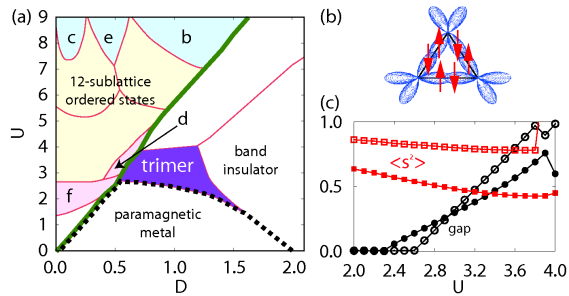}
\caption{(Color online). 
(a) Ground-state phase diagram 
for the multiorbital Hubburd model 
at the electron density $n=2$ obtained by the Hartree-Fock calculation.
The thick curve represents 
the boundary between magnetic and nonmagnetic solutions 
(the upper left is magnetic). 
The dotted curve represents the metal-insulator transition line. 
Spin and orbital patterns for the states b-f are drawn in Figs.~\ref{fig:mosikizu}(b)-(f), respectively.
(b) A schematic picture of the spin and orbital state in the trimer phase.
(c) Total spin $\langle S^2 \rangle$ (square) and energy gap (circle) as a function of $U$ at $D=0.7$ (open) and 1.1 (closed). 
}
\label{fig:sankakusouzu24}
\end{center}
\end{figure}

Figure~\ref{fig:sankakusouzu24}(a) 
shows the representative result for 
the ground-state phase diagram as functions of $U$ and $D$. 
We take $\eta=0.1$, i.e., $U^\prime = 0.8U$ and $J_{\rm H}=0.1U$.
When $U=0$, 
the system exhibits a metal-insulator transition 
from paramagnetic metal to 
band insulator at $D=2$ 
by the trigonal-field splitting of $a_{1g}$ and $e_g^\prime$ bands.
On the other hand, at $D=0$, the system is insulating for all $U>0$ 
because of the perfect nesting of the Fermi surface. 
In the large-$U$ region, we obtain the four-sublattice orbital-ordered ferrimagnetic state [Fig.~\ref{fig:mosikizu}(c)], 
in agreement with the above result for the large-$U$ effective model in Eq.~(\ref{eq:Heff}). 
We, however, do not find any trimer-type threefold ordering 
in the entire range of $U$ at $D=0$. 

When $D$ and $U$ become both finite, 
the system exhibits a variety of phases with different spin and orbital ordering~\cite{note}. 
Among them, the most interesting phase is the 
three-sublattice spin-orbital ordered state 
in the range of $0.5 \lesssim D \lesssim 1.5$ and $2 \lesssim U \lesssim 4$.
This phase is insulating 
[gap is plotted in Fig.~\ref{fig:sankakusouzu24}(c)]
and nonmagnetic, appearing
on the verge of the metal-insulator boundary and 
the magnetic-nonmagnetic boundary 
[dotted and thick curves in Fig.~\ref{fig:sankakusouzu24}(a), respectively]. 
In other words, 
it is stabilized in 
the competing region among the paramagnetic metal, band insulator, and 
magnetic Mott insulator. 

The three-sublattice ordered state shows a trimer-type orbital ordering, 
similar to that in the previously-proposed trimer state in Ref.~\citen{Pen1997}. 
The spin state is, however, quite different from the $S=1$ spin-singlet;  
each of the two $\sigma$-bond orbitals is dominantly occupied 
by up or down electron at each site, and 
the intersite $\sigma$ bond is formed by a pair of the spin-up and down orbitals, 
as shown in Fig.~\ref{fig:sankakusouzu24}(b).
This is different from the $S=1$ high-spin state with the total spin $\langle S^2 \rangle \simeq 2$ 
[see Fig.~\ref{fig:sankakusouzu24}(c)]. 

This state is distinct from that expected in 
the two well-known limits for the bond formation. 
One is a spin-Peierls type instability 
in the strong coupling picture, as discussed in Refs.~\citen{Onoda1991} and \citen{Pen1997}; 
in this case,
the spin-singlet bond is formed through the superexchange coupling between the localized $S=1$ spins. 
The other limit is the Peierls instability to form a bonding orbital in the weak coupling picture \cite{Peierls}; 
here each bonding orbital is unpolarized and spins do not play an important role. 
In our trimer state, the bonding orbitals are spin polarized 
to form the intersite spin-up and down pairs, 
whereas the polarization is cancelled at each site between two occupied orbitals. 
This is considered to be a compromise between the spin-Peierls-type superexchange physics 
and the Peierls-type bonding orbital formation.

\begin{figure}[t]
\begin{center}
\includegraphics[width=7.5cm,clip]{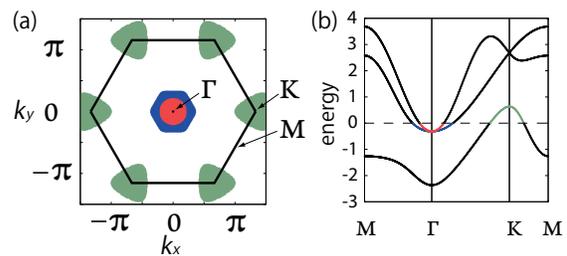}
\caption{(Color online). 
(a) Fermi surfaces at $(D,U) = (1, 2)$ obtained by the Hartree-Fock calculation. 
The large hexagon represents the first Brillouin zone. 
Dark grey circle and hexagon around the $\Gamma$ point represent 
electron pockets. 
The light grey triangles around the K points represent hole pockets. 
(b) Corresponding band dispersions. 
Fermi energy is set to be zero.}
\label{fig:bandstructure}
\end{center}
\end{figure}

The stabilization mechanism of the trimer-type orbital ordering 
is understood as follows. 
In the paramagnetic metal close to the metal-insulator phase boundary, 
one of the three bands makes a hole pocket of the Fermi surface 
around the K point, and the other two form 
two electron pockets around the $\Gamma$ point, 
as exemplified in Fig.~\ref{fig:bandstructure} for the case of $(D,U)=(1,2)$.  
As approaching 
the metal-insulator phase boundary, 
the pockets shrink and the nesting is developed 
between the hole pocket and electron pockets. 
At some point under sufficient nesting, 
finite Coulomb interactions induce an instability 
toward some symmetry breaking, 
as often seen in spin- or charge-density-wave transitions. 
In the present case, 
the nesting is between the different orbitals, and 
therefore, such instability gives rise to an orbital ordering. 
The ordering is of period three 
because the vector connecting the $\Gamma$ and K points is 
one third of the reciprocal lattice vector. 
Because this picture is simple and robust in the presence of the trigonal-field splitting, 
we believe that the nesting mechanism and the resulting trimer-type orbital order will survive 
even when going beyond the mean-field approximation and including fluctuation effects.

Let us discuss our results in comparison with experiments. 
First of all, our nonmagnetic trimer state appears in the intermediately correlated region, 
which is in 
qualitative agreement with experiments in LiVO$_2$. 
In particular, the fact that our trimer is in between the band insulator and Mott insulator 
well accounts for the comparable energy scale between the spin and charge gap. 
The calculated values of the gap, 
shown in Fig.~\ref{fig:sankakusouzu24}(c), 
is on the order of $0.1$~eV, when considering 
the bandwidth $\sim 4$~eV in the first-principles calculation~\cite{Ezhov1998}. 
This is compared favorably with the experimental estimate~\cite{Tian2004}.
Furthermore, our trimer state is on the border of the metal-insulator transition 
to the paramagnetic metal; this agrees with the experimental trend in 
the substituted materials LiV$X_2$ ($X$=O,~S,~Se)~\cite{Katayama2009}. 
Our trimer state is nonmagnetic, but each occupied orbital is spin polarized 
under the electron correlation, which is clearly distinct from 
the low-spin nonmagnetic band insulator under strong trigonal field: 
It is desired to reexamine the x-ray absorption spectra, 
which was argued to be consistent with the high-spin state~\cite{Pen1997B}, 
by taking our spin-orbital ordered trimer state into consideration. 
To confirm our scenario, it is important to identify the band structure and the Fermi surfaces 
in the metallic compounds, such as LiVSe$_2$, experimentally or by first-principles calculations. 
A quantitative estimate of Coulomb interactions is also helpful. 
It will be also interesting to experimentally examine the effect of uniaxial pressure 
which dominantly affects $D$.

Interestingly, our phase diagram includes the square-type AF state [Fig.~\ref{fig:mosikizu}(b)] 
in the region $U \gtrsim7$ for nonzero $D$. 
The spin-orbital order coincides with that in the lowest-temperature phase 
in a related compound NaVO$_2$~\cite{McQueen}. 
The compound is indeed more strongly-correlated than LiVO$_2$, 
evidenced by the larger charge gap and lattice constant. 
We, however, note that a LSDA+$U$ calculation predicted 
the opposite sign of $D$ for NaVO$_2$~\cite{Jia2009}. 
Extension of our analysis in a wide range of parameters including $D<0$ is left for future study.  

In our study, we considered the $\sigma$-bond overlap $t_{\sigma}$ only. 
We confirmed that our trimer state remains robust for more general band structure 
when $t_{\sigma}$ is dominant. 
We note that the value of $D$ for our trimer state 
is larger than that obtained by the first-principles calculation~\cite{Ezhov1998}. 
We also note that $U$ might be small compared to the estimate from 
fitting of the photoemission spectra for a related perovskite~\cite{Bocquet1996}. 
The parameter range for our trimer state, however, may be extended to smaller $D$ and larger $U$ region when we include the concomitant lattice distortions with trimer formation, as observed in experiments. 
We confirmed this tendency 
by considering the model (\ref{eq:fullHamiltonian0}) on an isolated three-site cluster.
Meanwhile, the cluster calculation predicts that
another magnetic trimer state (d) is also extended and reach $D=0$.
We speculate that this state is related with the spin-singlet trimer state
obtained in the exact diagonalization for a three-site cluster at $D=0$ in Ref.~\citen{Pen1997}. 
An interesting question is, in the presence of lattice modulation, 
how the magnetic and nonmagnetic trimer states are modified 
in more sophisticated calculations beyond the mean-field approximation. 
This problem is left for future study.

In conclusion, we have investigated the origin of the trimer formation in 
the Hubbard model with $t_{2g}$ orbital degeneracy on a triangular lattice. 
We found that a nonmagnetic trimer state with three-sublattice spin-orbital ordering 
is stabilized by nesting between different orbitals 
on the border of metal-insulator transition under the trigonal crystal field. 
This state takes the most efficient of both the spin-Peierls instability assisted by orbital ordering 
in the strong coupling picture and the Peierls instability by forming bonding orbitals 
in the weak coupling picture. 
Our result underlies a general mechanism of the cluster formation 
in the intermediately-correlated orbital degenerate systems. 
It also potentially provides comprehensive understanding of a variety of 
the ground states in the related compounds LiV$X_2$ ($X$=O,~S,~Se) and NaVO$_2$.

\begin{acknowledgment}

The authors thank D. Khomskii, H. Takagi, J. Matsuno, and N. Katayama for fruitful discussions.
This research was supported by KAKENHI (No. 19052008, 17740244, and 16GS0129), and 
Global COE Program ``the Physical Sciences Frontier." 
Y. M. acknowledges the hospitality of KITP Santa Barbara, 
where a part of this work was done. 
\end{acknowledgment}


\begin{thebibliography}{999}
\bibitem{note3} We can estimate the lower bound of $E_{\rm trimer}$ 
by decomposing the effective spin Hamiltonian into six-site clusters. 
The estimate is still higher than $E_{\rm ferri}$.
\bibitem{note} The appearance of 12-site orders 
implies that larger unit cell is necessary to discuss the correct ordering patterns 
in this intermediate-$U$ region. 
Other phases in the smaller $U$ region are basically understood 
from the nesting of the Fermi surface. 
The phase extending near $U \sim 4D$ for $U \gtrsim 4$ exhibits 
a four-sublattice spin-orbital order; there, each orbital has spin polarization, but the ploarization is cancelled at each site.
\end{thebibliography}
\end{document}